\documentclass[11pt,a4paper,notitlepage]{article}
\usepackage[utf8]{inputenc}
\usepackage[T1]{fontenc}
\usepackage{amsmath, amsthm, amssymb, amsfonts}
\usepackage{graphicx}
\usepackage{enumerate}
\usepackage[table]{xcolor}% http://ctan.org/pkg/xcolor
\usepackage{listings}
\usepackage{float}
\usepackage{subcaption}
\usepackage{booktabs}
\usepackage{bm}
\usepackage{nicefrac}
\usepackage{placeins}
\usepackage{natbib}
\usepackage{multirow}
\usepackage[title]{appendix}

\setcitestyle{authoryear, open={(},close={)}}

\usepackage{authblk}

\usepackage{color} %red, green, blue, yellow, cyan, magenta, black, white
\definecolor{mygreen}{RGB}{28,172,0} % color values Red, Green, Blue
\definecolor{mylilas}{RGB}{170,55,241}
\definecolor{MyDarkGreen}{rgb}{0.0,0.4,0.0}
\definecolor{Blue}{rgb}{0.0,0,0.4}
 \usepackage{hyperref}
 
 \usepackage[colorinlistoftodos]{todonotes}

\usepackage{geometry}
\usepackage{fancyhdr}
\newcommand{\thetab}{\bm{\theta}}

\newcommand{\xb}{\bm{x}}
\newcommand{\yb}{\bm{y}}

\newcommand{\Qb}{\bm{Q}}
\newcommand{\Gb}{\bm{G}}

\newcommand{\R}{\mathbb{R}}
\newcommand{\p}{\mathbb{P}}
\newcommand{\Q}{\mathbb{Q}}

\newcommand{\E}{\mathrm{E}}

\newcommand{\mub}{\bm{\mu}}

\newcommand{\Yb}{\bm{Y}}

\newcommand{\Ub}{\bm{U}}
\newcommand{\Vb}{\bm{V}}
\newcommand{\Ab}{\bm{A}}
\newcommand{\Sigmab}{\bm{\Sigma}}
\newcommand{\Xb}{\bm{X}}

\newcommand{\eb}{\bm{e}}
\newcommand{\fb}{\bm{f}}

\newcommand{\Cb}{\bm{C}}

\DeclareMathOperator{\MVN}{MVN}
\DeclareMathOperator{\CRPS}{CRPS}
\DeclareMathOperator{\SCRPS}{SCRPS}
\DeclareMathOperator{\rCRPS}{rCRPS}
\DeclareMathOperator{\sroot}{S_{root}}
\DeclareMathOperator{\slog}{S_{log}}

\DeclareMathOperator{\LOOS}{LOOS}

\usepackage{pifont}

\usepackage{soul}
\title{{\huge Fast and robust cross-validation-based scoring rule inference for spatial statistics}}

\author[1]{Helga Kristín Ólafsdóttir}
\author[1]{Holger Rootzén}
\author[2]{David Bolin}
\affil[1]{Department of Mathematical Sciences, Chalmers University of Technology and University of Gothenburg, Gothenburg, Sweden}
\affil[2]{Computer, Electrical and Mathematical Sciences and Engineering Division, King Abdullah University of Science and Technology, Thuwal, Saudi Arabia}
\date{}

\begin{document}
\maketitle

\begin{abstract}
Scoring rules are aimed at evaluation of the quality of predictions, but can also be used for estimation of parameters in statistical models. 
We propose estimating parameters of multivariate spatial models by maximising the average leave-one-out cross-validation score. 
This method, $\LOOS$, thus  optimises predictions instead of maximising the likelihood. The method allows for fast computations for Gaussian models with sparse precision matrices, such as spatial Markov models. 
It also makes it possible to tailor the estimator's robustness to outliers  and their sensitivity to spatial variations of uncertainty through the choice of the scoring rule which is used in the maximisation. 
The effects of the choice of  scoring rule which is used in $\LOOS$  are studied by simulation in terms of computation time, statistical efficiency, and robustness. 
Various popular scoring rules and a new scoring rule, the root score, are compared to maximum likelihood estimation. 
The results confirmed that for spatial Markov models the computation time for $\LOOS$  was much smaller than for maximum likelihood estimation. Furthermore, the standard deviations of parameter estimates were smaller for maximum likelihood estimation, although the differences often were small.  
The simulations also confirmed that the usage of a robust scoring rule results in robust LOOS estimates and that the robustness provides better predictive quality for spatial data with outliers. 
Finally, the new inference method was applied to ERA5 temperature reanalysis data for the contiguous United States and the average July temperature for the years 1940 to 2023, and this showed that the LOOS estimator provided parameter estimates that were more than a hundred times faster to compute compared to maximum-likelihood estimation, and resulted in a model with better predictive performance.

\end{abstract}

\section{Introduction}

Proper scoring rules are widely used for forecast evaluation and statistical model comparisons in areas such as meteorology, climate science and finance \citep{Lerch2013ComparisonForecasting,Friederichs2012,Carvalho2016AnRules,Serafini2022RankingEnvironment2}. The observations are compared to the  predictive distribution through scores, with higher values implying a better prediction. To keep the forecaster earnest, the expected score should be maximised for the true model. When this property holds, the score is said to be proper \citep{Gneiting2007}. 

Specifically, a scoring rule $S$ is a  function which takes a  predictive distribution $\p$ and an observation $x$ of a random variable $X$ and returns a real-valued score $S(\p,x)$ which gives a measure of the quality of the prediction, where in this paper higher values represent higher quality. Suppose $X$ has distribution $\Q$, and let $S(\p,\Q):=E_\Q[S(\p,X)]$ denote the expectation of  $S(\p,X)$. 
The scoring rule $S$ is said to be proper for a set of distributions $\mathcal{P}$ if for all $\p,\Q\in\mathcal{P}$
\begin{equation*}
    S(\p,\Q)\leq S(\Q,\Q).
\end{equation*}
 If strict inequality holds, the score is said to be strictly proper. 
 
In practise, average scoring rules are used for model comparison. That is, if $\yb=(y_1, \ldots, y_m )$ is a vector of observations, and $\p$ denotes a statistical model with corresponding predictive distributions, $\p_1, \ldots, \p_m$, then the score of $\p$ is typically computed as
\begin{equation*}
 \label{eq:averagescore}
   \bar{S}(\p,x) =  \frac{1}{m}\sum_{i=1}^m S(\p_i,y_i).
 \end{equation*}
It is easy to see that if $S$ is proper, then so is the average score.  

Many proper scoring rules exist, such as the logarithmic score ${\slog(\p, y) = \log f (y)}$, where $f(\cdot)$ is the density of $\p$~\citep{Good1952RationalDecisions}. Another popular scoring rule is the Hyvärinen score~\citep{Hyvarinen2005EstimationMatching} ${S(\p,y) = \partial^2 \log f(y) - 0.5 |\partial\log f(y)|^2}$, where $\partial$ is the derivative with respect to $y$. The arguably most popular scoring rule is the continuous ranked probability score (CRPS)~\citep{Baringhaus2004, Gneiting2007, Zamo2018}, which is a kernel score \citep{Dawid2007TheRules} and a part of a wider class of proper scoring rules called generalised kernel scores \citep{Bolin2023} that can be formulated as
\begin{equation}\label{eq:proper:bolin}
    S_g^h(\p,y)=h(\E_{\p\p}[g(X,X')])+2h'(\E_{\p\p}[g(X,X')])(\E_{\p}[g(X,y)]-\E_{\p\p}[g(X,X')])
\end{equation}
where $g$ is a non-negative continuous negative definite kernel and $h$ is a monotonically decreasing convex function~\citep{Bolin2023}. For CRPS, the function $h$ is chosen as $h(x)=-\frac{1}{2}x$ and $g(x,y)=|x-y|$, while $h(x)=-\log(x)$ and $g(x,y)=|x-y|$ results in the scaled CRPS (SCRPS)~\citep{Bolin2023}.  
The choice $h(x)=-\sqrt{x}$ and $g(x,y)=|x-y|$ results in a score which in this paper will be called the root score and denoted $\sroot$.  The properties of  $\sroot$
are in between those of $\CRPS$ and the $\SCRPS$ in terms of, e.g., robustness, see  Appendix \ref{app:sroot:scale:invariance}.

While all these scores are proper, they emphasise different aspects of the predictions and have various advantages and disadvantages. The Hyvärinen score is homogeneous in the density function, meaning that the normalising constant is unnecessary for parameter estimation, which can be beneficial when the normalising constant is difficult to calculate \citep{Hyvarinen2005EstimationMatching,Hyvarinen2007SomeMatching,DeFondeville2018}. The logarithmic score and the SCRPS are scale invariant, meaning that the observations will contribute equally to the expected score, regardless of the model uncertainty at the observed locations \citep{Bolin2023}. For further details on scale invariance, see Appendix \ref{app:sroot:scale:invariance}. Moreover, for models with unknown or ill-obtainable density, inference can be found through scores such as the CRPS, since it consists of expected values that can be computed using Markov Chain Monte Carlo (MCMC) methods.
Therefore, the choice of scoring rule is important when comparing statistical models. 

Proper scoring rules have also been used for statistical inference. For a parametric statistical model, with parameters $\thetab$, optimising a proper score $S$ with respect to the parameters yields an asymptotically unbiased estimator \citep{Dawid2016MinimumInference}, and performing statistical inference based on scoring rules can be advantageous. For example, \cite{DeFondeville2018} used scoring rules for inference of certain extreme value models which are difficult to fit using likelihood-based methods. As pointed out by \cite{Gneiting2007}, the advantage with optimum score estimation is that the choice of the scoring rule can be adapted to the problem at hand. One example is \citet{Gneiting2005CalibratedEstimation} where they showed empirically that a predictive regression model estimated using a CRPS-based optimum score estimator performed better than one based on maximum likelihood plug-in estimates for a meteorological problem. Optimising strictly proper scores forms a special case of $M$-estimation \citep{huber2011robust} as for example optimising the logarithmic score in one dimension corresponds to log-likelihood estimation. See \citet{Gneiting2007} and \citep{Dawid2016MinimumInference}, and the references therein, for further details about this. 

%For location parameter estimation on a Gaussian distribution with known variance the corresponding $M$-estimator is also known and discussed in \citet{Gneiting2007}.\todo[color=green]{add more examples here, and note the connection with M estimators (Gneiting 9.1)}

In this work, we focus on the estimation of random field models in spatial statistics through optimum score estimation. As far as we know, there has not been any systematic comparison between different scoring rules in this scenario, which is a main motivation for this work.
We propose a scoring rule inference approach based on maximising the average score of leave-one-out conditional predictions. Using this leave-one-out cross-validation ($\LOOS$) score  for estimation thus optimises the predictive ability of the model instead of maximising the likelihood. We show that this approach has two main advantages. First, for spatial Markov models, the computational cost is in general much lower than a standard likelihood approach.   Second, the inference method can be made robust against outliers by choosing a robust scoring rule. For this estimation approach, we compare different choices of scoring rules in terms of computation time, statistical efficiency and robustness. %\todo{describe in more detail}

The rest of the article is organised as follows. Section \ref{sec:estimation} introduces the details of the a leave-one-out cross-validation approach to model inference. Section \ref{sec:results:simulation} contains results from inference comparison using the different approaches on the spatial models in terms of time, accuracy, robustness and predictive quality. A case study on rainfall anomalies is carried out in Section \ref{sec:results:casestudy}. Finally, discussion and conclusions are found in Section \ref{sec:conclusion}. Details on the scoring rules and the models used for comparison are described in two appendices.

\section{Scoring rule inference}\label{sec:estimation}
A straightforward approach to scoring rule parameter estimation for multidimensional random variables is to use multidimensional scoring rules, such as the log-likelihood score or energy scores \citet{Gneiting2008}. However, in spatial statistics the aim often is spatial prediction. Different models (or prediction methods) are then compared by cross-validation: a set of values in a spatial data set are in turn removed from the data, the remaining data are used to find the predictive distribution for the left out values, and finally the value of the scoring rule used on this prediction is computed. The sum of these scores over all values in the data set are then used to compare different prediction methods.
In this section, we introduce and use leave-one-out score (LOOS) optimisation for parameter estimation. We give the general background of maximum score estimation, the formal definition of LOOS, examine the computational requirements for LOOS in different scenarios and compare these with full likelihood estimation, and discuss robustness of the LOOS estimators. 

\subsection{Background}\label{sec:background}
Consider observations of a random variable $X\sim\Q$. 
%\cite{Dawid2016MinimumInference} use the score for inference. 
For a proper scoring rule $S$, the divergence $D$ (or discrepancy function) associated with score $S$ is
\begin{equation*}
D(\p,\Q)=S(\Q,\Q)-S(\p,\Q),
\end{equation*}
where $S(\Q,\p)$ is the expected score of forecast $\p$, and $H(\Q)=S(\Q,\Q)$ is the generalised entropy 
%\todo{reference}
associated with $S$. Now assume that $\Q$ belongs to a family of parametric models $P_{\thetab}$ with  true parameter $\thetab_0$. Then under conditions discussed in \cite{Dawid2016MinimumInference} we have that
\begin{equation*}\label{eq:min:divergence}
    \thetab_0=\arg\min_{\thetab} D(\p_\theta,\Q)=\arg\max_{\thetab} S(\p_{\thetab},\Q),
\end{equation*}
where the second equality is because the generalised entropy does not depend on $\thetab$.

For $m$ independent observations $y_1,y_2,...,y_m$ of $X$, the average score of the observations is used for inference, and the estimator becomes
\begin{equation}\label{eq:score:estimator}
    \hat{\thetab}=\arg\max_{\thetab} \frac{1}{m}\sum_{k=1}^m S(\p_{\thetab},y_k).
\end{equation}
%    \todoin{(or in our case, the max, since we have positively oriented scores).}
%
According to \citet{Dawid2016MinimumInference}, again under suitable conditions, the estimation equation
\begin{equation*}
  \nabla_{\thetab} \hspace{2mm} \frac{1}{m}\sum_{k=1}^m S(\p_{\thetab},y_k) = 0
\end{equation*}
used to solve Eq.~\eqref{eq:score:estimator} is unbiased.
%for every proper scoring rule $S$. 
Moreover, under suitable regularity conditions (for references see \citet{Dawid2016MinimumInference}), the estimator in Eq.~\eqref{eq:score:estimator} is consistent and asymptotically normal, with mean $\thetab_0$ and variance $V$, where $V^{-1}$ is the Godambe information matrix~\citep{Godambe1960}, i.e.
%\todo{compare to theoretical values?}Theorem 4.1. in~\citep{Dawid2016MinimumInference} states that under suitable regularity conditions, the scoring rule estimator in Eq. \eqref{eq:score:estimator} is consistent and asymptotically normal with mean $\theta_Q$ and variance $V$, where
\begin{equation}\label{eq:godambe:v}
    V=K^{-1}J(K^{-1})^\top,
\end{equation}
with
\begin{equation*}\label{eq:godambe:j:k}
    \begin{aligned}
        J&=\E_{\p_{\thetab_0}}[\nabla_{\thetab} S(\p_{\thetab},X)|_{\thetab=\thetab_0}\nabla_{\thetab} S(\p_{\thetab},X)^\top|_{\thetab=\thetab_0}]\\
        K&=\E_{\p_{\thetab_0}}\left[\nabla^2_{\thetab} S(\p_{\thetab},X)|_{\thetab=\thetab_0}\right].
    \end{aligned}
\end{equation*}
% \begin{equation*}
%     \begin{aligned}
%         J&=E_Q[s(\theta_Q)s(\theta_Q)^T]\\
%         K&=E_Q\left[\frac{\partial s(\theta)}{\partial\theta^T}\right].
%     \end{aligned}
% \end{equation*}
% and
% \begin{equation*}
%   s(x,\theta)=\nabla_\theta S(\p_\theta,x)=\dfrac{\partial S(\p_\theta,x)}{\partial \theta}.  
% \end{equation*}
When $S$ is the logarithmic score, $V^{-1}$ is the Fisher information matrix.

\subsection{Leave-one-out scoring rule parameter estimation for spatial statistics}\label{sec:spatial:inference}

 As set out above, we propose using the leave-one-out score for parameter estimation. Suppose that we have a set of data $y_1,\ldots, y_n$ which is an observation of a multivariate random variable $\Xb = (X_1,\ldots, X_n)$ with distribution $\p(\thetab)$ depending on a parameter vector $\thetab \in \mathbb{R}^p, p\geq 1$. Let $\p_{i|-i}$ denote the conditional distribution of $X_i|\Xb_{-i}$, where $\Xb_{-i}$ denotes all elements of $\Xb$ except for $X_i$. For observations $\yb$ of $\Xb$, we condition each $\p_{i|-i}$ on $\Xb_{-i}=\yb_{-i}$ and define the leave-one-out score as the average of the one-dimensional scores over these conditional distributions,
\begin{equation*}\label{eq:score:sum:sii}
    LOOS(\p_{\thetab},\yb)=\frac{1}{n}\sum_{i=1}^n S(\p_{i|-i}(\thetab),y_i).
\end{equation*} 
An estimate of ${\thetab}$ then is obtained as the maximiser of this scoring rule, 
\begin{equation}\label{eq:score:estimator:conditional}
    \hat{\thetab} = \arg\max_{\thetab} \frac{1}{n}\sum_{i=1}^n S(\p_{i|-i}(\thetab),y_i).
\end{equation}
The composite $\LOOS$ score is clearly a proper scoring rule as long as $S$ is. This is thus a special case of the general procedure in Section~\ref{sec:background} where the scoring rule is chosen as the composite $\LOOS$ score. If $S$ is chosen as the logarithmic score, the $\LOOS$ score is nothing but the pseudo-likelihood of \citet{Besag1975StatisticalData}, which also was pointed out by 
\cite{Dawid2016MinimumInference} Example 1 in the context of Markov models. 

The LOOS estimator is thus easy to maximise numerically for any model where $S(\p_{i|-i}({\thetab}),\Xb_i)$ can be evaluated analytically. In the following subsections, we illustrate this for Gaussian random fields, and the analytic expressions for several common scoring rules for Gaussian distributions are given in Appendix \ref{app:normal:score}. 

The efficiency of the estimator can be assessed through the Godambe information matrix in Eq.~\eqref{eq:godambe:v} with 
\begin{equation*}
    \begin{aligned}
        J&=\frac{1}{n}\sum_{i=1}^n\E_{\p_{\thetab_0}}[\nabla_{\thetab}  S(\p_{i|-i}({\thetab}),X_i)|_{\thetab=\thetab_0}\nabla_{\thetab}  S(\p_{i|-i}(\thetab),X_j)^\top|_{\thetab=\thetab_0}]\\
        K&=\frac{1}{n}\sum_{i=1}^n \E_{\p_{\thetab_0}}\left[\nabla^2_{\thetab} S(\p_{i|-i}(\thetab),X_i)|_{\thetab=\thetab_0}\right].
    \end{aligned}
\end{equation*}
Thus, the asymptotic efficiency clearly depends on the choice of scoring rule $S$, and in the following we will investigate this numerically in a few common scenarios. 

\subsection{LOOS estimation for Gaussian random fields}
A standard scenario in spatial statistics is that the data are modelled as observations of a Gaussian random field $X$. Let $X_i = X(s_i)$ be an observation of the field at location $s_i$. Then $\Xb=(X_1,...,X_n)\sim \MVN(\mub(\thetab),\Sigmab(\thetab))$ be a multivariate normal distribution with mean $\mub(\thetab)$ and covariance matrix $\Sigmab(\thetab)$ determined by the mean and covariance functions of $X$. In this case, 
$$
X_i|\Xb_{-i}=\xb_{-i} \sim N(\mu_i+\Sigmab_{i,-i}\Sigmab_{-i,-i}^{-1}(\xb_{-i}-\mub_{-i}),\Sigmab_{i,i} -\Sigmab_{i,-i}\Sigmab_{-i,-i}^{-1}\Sigmab_{-i,i}),
$$
where $\Sigmab_{-i,\cdot}$ and $\Sigmab_{\cdot,-i}$ denote all but the $i$-th row and column of $\Sigmab$, respectively, and $\Sigmab_{-i,i}$ denotes the $i$-th column of  $\Sigmab$ where the $i$-th element has been removed.
The $\LOOS$ expression can thus be calculated analytically for any scoring rule that has a closed-form expression for Gaussian distributions, see  Appendix~\ref{app:normal:score}.

Note that the computation of the conditional distribution requires the inverse of an $n-1\times n-1$ matrix, which for a general covariance matrix, $\Sigmab$, requires $O(n^3)$ floating point operations. As this has to be done $n$ times, it thus seems as if the total cost is $O(n^4)$, which is significantly higher than the $O(n^3)$ cost of evaluating the likelihood. However, the cost can be reduced significantly by first computing $\Sigmab^{-1}$ at $O(n^3)$ cost, and then obtaining all inverses $\Sigmab_{-i,-i}, i=1,\ldots n$ as rank-two updates of $\Sigmab^{-1}$. Specifically, let $\eb_i$ denote the column vector of length $n$ with all zero elements except a $1$ at the $i$-th entry, and let $\fb_i$ denote $i$-th column of $\Sigmab$ except that the $i$-th element of this vector is set to zero. Defining the matrices 
$
\Ub = \begin{bmatrix}
\eb_i & \fb_i
\end{bmatrix}$ and 
$\Vb = \begin{bmatrix}
\fb_i & \eb_i
\end{bmatrix}^{\top}
$
we then have via the Woodbury matrix identity that 
\begin{equation*}
    (\Sigmab-\Ub\Vb)^{-1} = \Sigmab^{-1}-\Sigmab^{-1}\Ub(I-\Vb\Sigmab^{-1}\Ub)^{-1}\Vb\Sigmab^{-1},
\end{equation*}
which only requires the inverse of a $2\times2$ matrix. Omitting the $i$-th row and $i$-th column from $(\Sigmab-\Ub\Vb)^{-1}$ yields $\Sigmab_{-i,-i}^{-1}$. Thus, the computational cost of LOOS estimation is of the same order as the cost of likelihood-based estimation in this case. However, as seen below, in important special cases LOOS is faster.

\subsection{LOOS estimation for Gaussian Markov random fields}\label{sec:gmrftime}
In spatial statistics, Gaussian Markov random fields (GMRFs) are often used to reduce computational costs. Examples include conditional autoregressive (CAR) models \citep{Besag1975StatisticalData} for areal data and the stochastic partial differential equation (SPDE) approximations of Gaussian random fields with Mat\'ern covariances \citep{Lindgren2011}. 

A GMRF is a multivariate normal random variable $\Xb=(X_1,...,X_n)\sim \MVN(\mub(\thetab),Q(\thetab)^{-1})$, where the sparsity pattern of the precision matrix $\Qb=\Sigmab^{-1}$ determines conditional independence properties of the field \citep{Rue2005}. Typically, $\Qb$ is a highly sparse matrix, and this sparsity can be used to reduce the computational costs for likelihood-based inference down to $O(n)$ in one spatial dimension, $O(n^{3/2})$ in two spatial dimensions, and $O(n^{2})$ in three spatial dimensions \citep{Rue2005}. 

 The fact that the model is specified through the precision matrix also simplifies LOOS estimation. The reason is that the conditional distribution in terms of the precision matrix is 
\begin{equation*}
X_i|\Xb_{-i}=\yb_{-i} \sim N(\mu_i-\frac{1}{Q_{ii}}\sum_{j\neq i}Q_{ij}(y_j-\mu_j), Q_{ii}^{-1}).
\end{equation*}

Thus, no matrix inversion is required to compute the conditional variance, and the conditional mean can be computed as 
\begin{equation*}\label{eq:smart:solution}
    \mu_i-\frac{1}{Q_{ii}}\sum_{j\neq i}Q_{ij}(y_j-\mu_j) = 
  \frac{1}{Q_{ii}}(\Qb\mub)_i-\frac{1}{Q_{ii}}(\Qb\yb)_i+y_i,
\end{equation*}
which means that the matrix vector product $\Qb\mub$ only has to be computed once, and which has cost $O(n)$ if $\Qb$ is sparse and only has a few non-zero elements per row.  
Thus, in total the computational cost for LOOS estimation for GMRFs is $O(n)$ in any spatial dimension, compared to the $O(n^{3/2})$ cost for the likelihood-based inference in two dimensions and $O(n^{2})$ in three dimensions. 

\subsection{LOOS estimation for latent GMRFs}
A common scenario is that the data $\yb$ is modelled as an observation of a GMRF ${\Xb \sim N(\mub, \Qb^{-1})}$ observed under Gaussian measurement noise, $\Yb|(\Xb=\xb) \sim N(\Ab\xb,\sigma_\varepsilon^2 I)$, 
where $\Ab\in \R^{m}\times \R^{n}$ is a projection matrix linking the GMRF to the observations. It is well-known \citep[see, e.g.][]{Rue2005} that
$\Xb|(\Yb=\yb)\sim N(\mub_{x|y},\Qb_{x|y}^{-1})$,
where $\Qb_{x|y} = \Ab^{\top}\Qb_\varepsilon \Ab+\Qb$ and $\mub_{x|y} = \Qb_{x|y}^{-1}\Ab^{\top}\yb/\sigma_\varepsilon^2$.
To perform the LOOS estimation, we need to compute the distribution of $Y_i|(\Yb_{-i}=\yb_{-i})$, which is a Gaussian distribution with mean 
$$
\mu_{i|-i} = \widehat{\Qb}_{i|-i}^{-1}\Ab_{-i,\cdot}^\top\yb_{-i}/\sigma_\varepsilon^2
$$
and variance $\sigma_\varepsilon^2 + q_i$, where $q_i$ is the $i$-th diagonal element of 
$$
\widehat{\Qb}_{i|-i} = \Ab_{-i,\cdot}^\top \Ab_{-i,\cdot}/\sigma_\varepsilon^2+\Qb
$$
Computing one of these conditional distributions has the same cost as computing the log-likelihood of the model. However, using the same type of low-rank updates as in the standard Gaussian field case, the computational cost of the LOOS expression can be reduced to be of the same order as the cost of evaluating the likelihood. 

\subsection{Robustness}
It is well-known that maximum likelihood estimation can be sensitive to outliers in the data. One popular measure of sensitivity is through the influence function (IF) \citep{Hampelincluencefunc}, which measures the standardised effect of a small contamination in the data at a point $y$. Let $\hat{\thetab}(\p)$ denote the $\LOOS$ estimator of Eq.~\eqref{eq:score:estimator:conditional} when the data comes from the model $\p$. The IF is then defined as 
$$
IF(\yb) = \lim_{\varepsilon\rightarrow 0} \frac{\hat{\thetab}(\p_\varepsilon) - \hat{\thetab}(\p_{\thetab})}{\varepsilon},
$$
where $\p_\varepsilon = (1-\varepsilon)\p_{\thetab} + \varepsilon \delta_{\yb}$. A standard definition of robustness is that the IF should be bounded. According to \cite[Eq. (27)]{Dawid2016MinimumInference}, the IF for $\LOOS$ can be rewritten as 
$$
IF(\yb) = K^{-1}\nabla_{\thetab} LOOS(\p_{\thetab},\yb) =  \frac{1}{n}\sum_{i=1}^n K^{-1}\nabla_{\thetab} S(\p_{i|-i}(\thetab),y_i).
$$
Thus, if the scoring rule $S$ is bounded in $y$ for all values of $\thetab$, then the $\LOOS$ estimator is robust. This boundedness of the scoring rule is in fact the definition of robustness of scoring rules used in \cite{Bolin2023}. \cite{Bolin2023} further introduced the notion of model sensitivity of a scoring rule as follows: If for a given $\p$ there exists a number $\alpha_{\p}$ such that $|S(\p, y)| \asymp |y|^{\alpha_{\p}}$, then $S$ is said to have model sensitivity $\alpha_{\p}$ for $\p$. They further introduced the sensitivity index $\alpha = \sup_{\p \in \mathcal{P}} \alpha_{\p}$ and noted that a scoring rule is robust if $\alpha = 0$. 

The model sensitivity of a scoring rule is useful as this provides a means to compare the sensitivity of different non-robust scoring rules, where a higher $\alpha_{\p}$ indicates a more sensitive scoring rule. Examples of the sensitivity index for Gaussian models for different scoring rules are shown in Table~\ref{tab:summaryscores}. The sensitivity indices of all scoring rules except for the root score $\sroot$ are derived in \cite{Bolin2023}. In the table, $\rCRPS$ refers to the robust CRPS, obtained by taking $g(x,y)=|x-y|1\{|x-y|<c\}$ for some $c>0$ and $h(x)=-\frac{1}{2}x$ \citep{Bolin2023}. This scoring rule is interesting from a robustness point of view as it has sensitivity index 0, and thus results in a bounded IF for the LOOS estimator. 

As discussed in the introduction, the root score, $\sroot$, scoring rule is obtained by choosing $h(x)=-\sqrt{x}$ and $g(x,y)=|x-y|$ in Eq.~\eqref{eq:proper:bolin}. This results in the score 
\begin{equation*}\label{eq:sroot}
    \sroot(\p,y)=-\frac{\E_{\p}[|X-y|]}{\sqrt{\E_{\p\p}[|X-Y|]}},
\end{equation*}
which can also be formulated in terms of CRPS as
\begin{equation*}\label{eq:sroot:crps}
    \sroot(\p,y) = \frac{1}{\sqrt{2}}\left(\frac{\CRPS(\p,y)}{\sqrt{|\CRPS(\p,\p)|}}-\sqrt{4|\CRPS(\p,\p)|}\right)
\end{equation*}
since $\CRPS(\p,y)=\frac{1}{2}\E_{\p\p}[|X-X'|]-\E_{\p}[|X-y|]$. The root score is interesting since it has the same sensitivity index as the SCRPS and CRPS and it has scale invariance properties which are in between those of $\CRPS$ and the $\SCRPS$, see  Appendix \ref{app:sroot:scale:invariance} for details and for the explanation of the scale dependence on $\sigma$ in Table~\ref{tab:summaryscores}. 

The values in the table suggest that only the rCRPS is robust, and that the CRPS is less sensitive to outliers than the log-score, as also noted by \cite{Gneiting2007}, but that neither the CRPS or the log-score are  robust. Further, neither the SCRPS or the root score are robust, and will therefore not result in a bounded IF for the $\LOOS$ estimator. 

\begin{table}[t]
\caption{Local scale invariance and sensitivity index of scoring rules for Gaussian models.}
    \label{tab:summaryscores}
    \centering
    \begin{tabular}{rcccc}
    \toprule
       Scoring rule & Sensitivity index & Robust & Scale dependence on $\sigma$ & Scale invariant \\
       \hline 
       log-score  & 2 & No & $\nicefrac1{\sigma^{2}}$ & Yes \\
       SCRPS      & 1 & No & $\nicefrac1{\sigma^{2}}$ & Yes \\
       Root-score, $\sroot$ & 1 & No & $\nicefrac1{\sigma^{1.5}}$ & No \\
       CRPS       & 1 & No & $\nicefrac1{\sigma}$ & No \\
       rCRPS      & 0 & Yes & $\nicefrac1{\sigma}$ & No \\
    \bottomrule
    \end{tabular}  
\end{table}

\section{Simulation studies}\label{sec:results:simulation}
In this section, we investigate the properties of $\LOOS$ estimation through simulation in a few different scenarios. First, we compare the $\LOOS$ estimator with maximum likelihood estimation using the different scoring rules shown in Table~\ref{tab:summaryscores} on a spatial GMRF model. Next, we consider a latent GMRF model, and a latent non-stationary GMRF model. Finally, we investigate the effects of the different estimation methods on predictive performance.

\subsection{Analysis of a GMRF model}\label{sec:results:gmrf}

\begin{figure}[t]
\centering
 \includegraphics[width=0.7\textwidth]{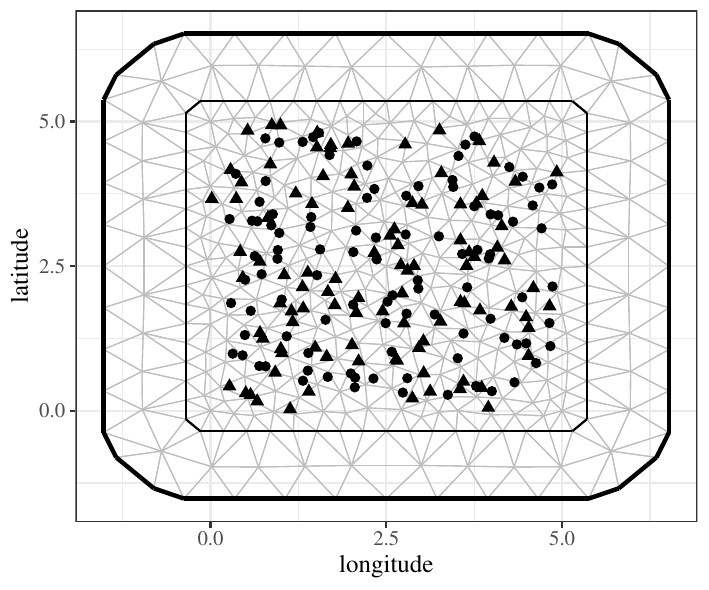}   
    \caption{Triangulation used for the direct and latent GMRF models as well as the observation locations for the latent GMRF models during training (circles) and testing (triangles).} \label{fig:spdemesh}
\end{figure}

The most common covariance model for Gaussian processes on $\mathbb{R}^d$ in spatial statistics is the Mat\'ern covariance 
\begin{equation*}\label{eq:gmrf:covariance}
    C(h) = \frac{\sigma^2}{2^{\nu-1}\Gamma(\nu)}(\kappa h)^\nu K_\nu(\kappa h),
\end{equation*}
where $\nu>0$ is a smoothness parameter, $\kappa>0$ represents the correlation range, and $\sigma^2$ is the variance of $\Xb$. A Gaussian process $\Xb$ with this covariance function can be represented as the solution to the stochastic partial differential equation (SPDE)
\begin{equation}\label{eq:spde}
    (\kappa^2-\Delta)^{\alpha/2}(\tau \Xb)=\mathcal{W},
\end{equation}
where $\Delta$ is the Laplace operator, $\Gamma$ is the gamma function, $K_\nu$ is the Bessel function of the second kind, $\tau^2=\Gamma(\nu)/(\sigma^2\kappa^{2\nu}(4\pi)^{d/2}\Gamma(\nu+d/2))$, $\alpha = \nu + d/2$, and $\mathcal{W}$ is a Gaussian white noise~\citep{Lindgren2011}. 

We fix $\alpha=2$ and use the SPDE approach of \citep{Lindgren2011} to obtain a GMRF approximation of this Gaussian process. A basis of piecewise linear basis functions $\{\varphi_i\}_{i=1}^{n}$ is constructed based on a  triangulation of the region of interest (Figure~\ref{fig:spdemesh}), where $\varphi_i$ is constructed to be 1 at the $i$-th node in the triangulation and zero at all other nodes. The mesh was obtained using the \texttt{inla.mesh.2d} function in R-INLA \citep{Rue2009} and has $n=435$ nodes. 
The Gaussian process is approximated as $x(s) = \sum_{i=1}^n x_i \varphi_i(s)$, where the vector of weights $\Xb$ is a GMRF, $\Xb \sim N(0,\Qb^{-1})$, with precision matrix 
\begin{equation}\label{eq:spdeQ}
\Qb = \tau^2(\kappa^2 \Cb + \Gb) \Cb^{-1} (\kappa^2 \Cb + \Gb).     
\end{equation}
Here $\Gb$ and $\Cb$ are two sparse matrices that only depend on the mesh, and not on the parameters of the model \citep{Lindgren2011}. 
The data for this model is $y_i = x_i, i=1,\ldots,n$ and the parameters to be estimated are $\thetab=(\tau,\kappa)$.

We simulate $10$ independent samples of the field, each with $435$ observations using the parameters $\thetab=(0.16,1.75)$, corresponding to a field with variance one and practical correlation range ($\sqrt{8}/\kappa$) approximately equal to 1.6. These parameters will be used in the following sections as well. For the rCRPS, we set $c=2$. 
Based on the data, the parameters are estimated using maximum likelihood and the $\LOOS$ estimator based on the log-score, CRPS, SCRPS, rCRPS and the $\sroot$ scoring rule. The optimisation was done using \texttt{optim} in R~\citep{Rsoftware}, and was repeated 300 times, each time with a new simulated data set. To test the sensitivity to outliers in the GMRF model, outliers were introduced by choosing an outlier's location $i$ at random from $1,..,m$ and then replacing $y_i$ by $|y_i|+5$. This is done independently for $k$ of the 10 samples generated of the field, for $k\in\{0,5,10\}$ and for each of the 300 independent repetitions.

\begin{figure}[t]
    \centering
    \includegraphics[width=0.9\textwidth]{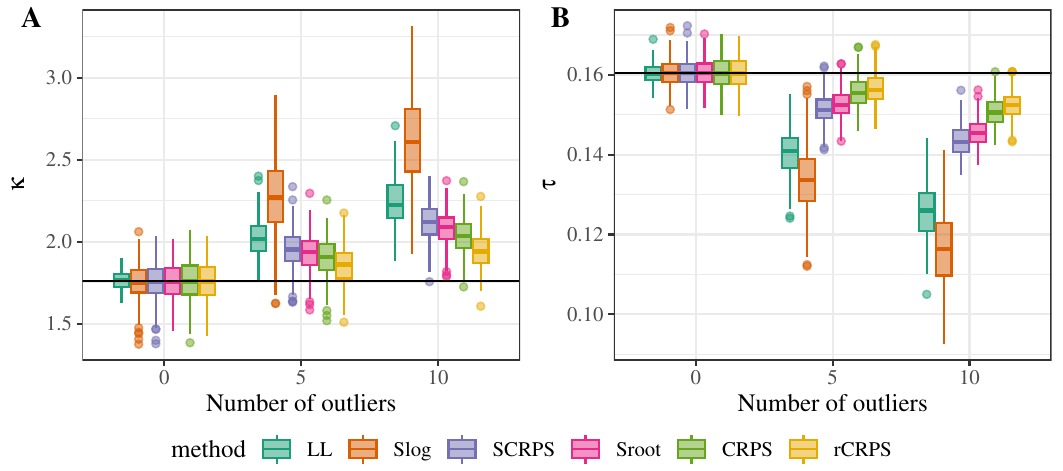}
    \caption{Estimated parameters for the direct GMRF model with no, five, and ten outliers, at most one in each independent observation.}
    \label{fig:repeated:estimation:m2}
\end{figure}

The results can be seen in Figure \ref{fig:repeated:estimation:m2}.  Without outliers, the estimates of all estimation methods centre around the true values, as expected as both the maximum-likelihood estimator and the various LOOS estimators are asymptotically unbiased. The maximum-likelihood estimator has lower variance than the LOOS estimators, which also is expected. The variances can be compared with the asymptotic variances of the estimators based on the Godambe information matrices. For the LOOS, the Godambe matrix was computed numerically by averaging the gradient and the hessian of the scores over 1000 simulated sets of observations. The results are shown in Table~\ref{tab:godambe}, which also shows the asymptotic standard deviations of the parameters for the case where the marginal standard deviation is $0.5$ and the practical correlation range is 1. 
%One parameter was varied at a time while the other parameter was kept fixed, as shown in Figure \ref{fig:godambe:m2}. \todo[color=green]{info about Godambe computations}%\todo[color=green]{Add log-score in figure}
One can note that the logarthmic score and the SCRPS have the lowest variance among the LOOS estimators, whereas the CRPS and the rCRPS have the highest in general.

\begin{table}[t]
\caption{Asymptotic standard deviations of the parameter estimates for the direct GMRF model according to the Godambe information matrices for the different estimation methods. The case $\thetab = (0.16, 1.75)$ is the one used in Figure~\ref{fig:repeated:estimation:m2} whereas the case $\thetab = (-1.6, 1.04)$ corresponds to a marginal standard deviation of $0.5$ and a practical correlation range of $1$.}
    \label{tab:godambe}
    \centering
    \begin{tabular}{rrcccccc}
    \toprule
     True parameters &  & LL & Slog & SCRPS & Sroot & CRPS & rCRPS  \\
       \hline 
      \multirow{ 2}{*}{$\thetab = (0.16, 1.75)$} & std $\log\tau$ & 0.444 & 0.469 & 0.643 & 1.33 & 2.2 & 3.39  \\
      & std $\log\kappa$ & 0.185 & 0.137 & 0.406 & 4.93 & 0.811 & 1.69 \\
       \hline 
      \multirow{ 2}{*}{$\thetab = (-1.6, 1.04)$} & std $\log\tau$ & 0.364 & 0.398 & 0.517 & 0.821 & 2.17 & 2.65  \\
      & std $\log\kappa$ & 0.262 & 0.253 & 0.522 & 1.29 & 6.93 & 7.64 \\
    \bottomrule
    \end{tabular}  
\end{table}

%\begin{figure}[t]
%    \centering    %\includegraphics{images/godambe/godambe_res_m220240630_103127_lines_dark2.pdf}
    %\includegraphics{images/godambe/godambe_res_m220240703_121340_lines.pdf}
    %\caption{Asymptotic variance for the direct GMRF model using the inverse Godambe matrix with $\log\tau = 0.5$ when varying $\kappa$ (top) and $\log\kappa = -1.8$ when varying $\tau$ (bottom).}
%    \label{fig:godambe:m2}
%\end{figure}

With growing number of outliers, the estimates diverge from the true parameter values and their variances increase. The log-likelihood and the $\LOOS$ with log-score are more sensitive to the outliers than the other scores, as expected from Table~\ref{tab:summaryscores}. The rCRPS is most robust, which is also expected since this is a robust scoring rule. The robustness of $\sroot$ lies between the SCRPS and the CRPS. The conditional scoring inference with the logarithmic score is the most sensitive. Thus, as expected, there is a tradeoff between low variance and robustness, where the rCRPS provides the most robust LOOS estimator but which also has the highest asymptotic variance.

\begin{table}[t]
\caption{Mean estimation runtime of the GMRF model using log-likelihood, and $\LOOS$   for no outliers, one outlier in half of the 10 samples (5 outliers) and one outlier
in each of the 10 samples. Standard deviation is reported within parentheses.}
    \label{tab:GMRF:time:hist}
    \centering
    \begin{tabular}{rccc}
    \toprule
       Scoring rule & 0 outliers & 5 outliers & 10 outliers \\
       \hline 
       Log-likelihood & 1.62  (0.375)&1.56  (0.339)&1.53  (0.328 )\\
      $\LOOS$, $\slog$& 0.575 (0.288)&  0.372 (0.178 )&0.375 (0.200)\\
       $\LOOS$, SCRPS      & 0.588 (0.235)& 0.483 (0.216)&0.421 (0.174)\\
       $\LOOS$, $\sroot$ & 0.567 (0.277)& 0.493 (0.218)&0.425 (0.176)\\
       $\LOOS$, CRPS       & 0.477 (0.200) & 0.458 (0.204)&0.444 (0.202)\\
       $\LOOS$, rCRPS      & 0.806 (0.219)& 0.786 (0.217)& 0.757 (0.221)\\
    \bottomrule
    \end{tabular}  
\end{table}

\begin{figure}[t]
    \centering  
    \includegraphics[width=0.6\linewidth]{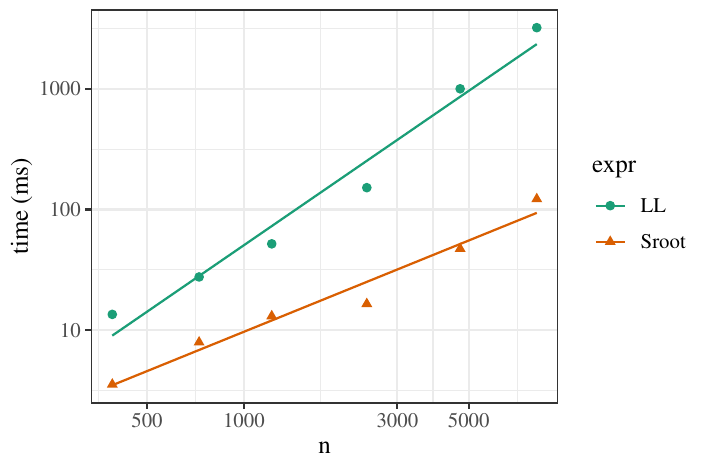}
    \caption{Average runtime as a function of the dimension of the GMRF, $n$, using the same parameters of the field but varying the number of nodes in the mesh.}
    \label{fig:runtimes}
\end{figure}

The runtime for the optimisation is shown in Table~\ref{tab:GMRF:time:hist}, and as expected from the results in Section~\ref{sec:gmrftime}, the runtime is lower for the LOOS inference than for the likelihood-based inference. It can also be noted that the various LOOS estimators have similar runtimes, except the rCRPS-based estimator which has a slightly longer runtime than the others.
To further illustrate the lower cost of the LOOS estimator, Figure~\ref{fig:runtimes} shows the average runtime for the maximum likelihood estimator and the LOOS $\sroot$ estimator as a functions of the dimension of the GMRF, $n$, using the same parameters of the field but varying the number of nodes in the mesh. As can be seen in the figure, the LOOS estimator has better scaling with $n$, which means that it is applicable to much larger datasets. 

\subsection{A latent GMRF model}\label{sec:latentGMRF}
As previously mentioned, a common scenario in spatial statistics is that the data is modelled as a latent GMRF observed under Gaussian measurement noise. To study the performance of the estimators in this scenario, we now assume that we have observations $y_1,\ldots, y_m$ at spatial locations $s_1,\ldots, s_m$ in the domain of interest (Figure~\ref{fig:spdemesh}). The data is modelled as ${\Yb|\Xb \sim N(\Ab\Xb,\sigma_\varepsilon^2 \mathbf{I})}$, where $\Ab$ is an $m \times n$ is a projection matrix linking the GMRF to the observations, with elements $A_{ij} = \varphi_j(s_i)$ and $\Xb \sim N(\mathbf{0}, \Qb^{-1})$ has precision as in Eq.~\eqref{eq:spdeQ}. 

We simulate 10 independent replicates of the field using the parameters $\thetab = (\sigma_\varepsilon,\tau,\kappa)$ with $(\tau,\kappa) =(0.16, 1.75)$ as before, and  $\sigma_\varepsilon=0.5$. %(0.1,0.5,1.75)$.
Based on the data, the parameters are again estimated using maximum likelihood and using the $\LOOS$ estimator based on the logarithmic score, CRPS, SCRPS, rCRPS and the $\sroot$ scoring rule. This was repeated $300$ times, each time with a new simulated data set. 
To study the sensitivity to outliers in this case, outliers were introduced by choosing a random location $i$ from $i=1,...,m$, and replacing $y_i$ by $|y_i|+5$. This was repeated for each of the $300$ data sets.

\begin{figure}
     \centering
     \includegraphics[width=\textwidth]{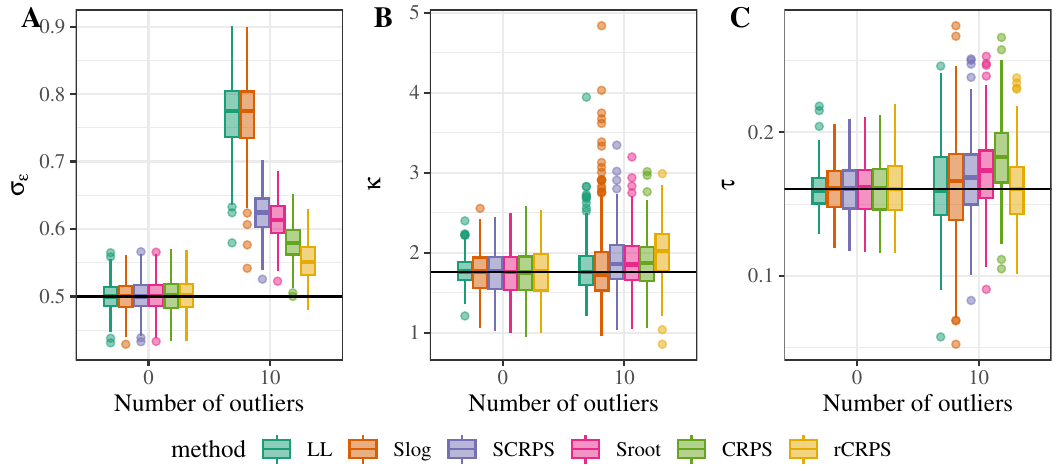}
     \caption{Estimated parameters for the latent GMRF model with and without outliers.}\label{fig:est:outlier:m3}
\end{figure}

All results can be seen in Figure~\ref{fig:est:outlier:m3}. The results are similar to those from the previous section, where the log-likelihood and the log-score are sensitive to outliers but have the lowest standard deviations of the parameter estimates when there are no outliers. 
The LOOS estimator based on the rCRPS is again most robust but also has the highest variance when there are no outliers.

The runtimes for the estimation are shown in Table \ref{tab:latent:GMRF:time:hist},  and as expected it is similar for all estimation methods. 

% \begin{figure}[h]
%     \centering
%     \includegraphics[width=0.8\textwidth]{images/GMRF/inlatimehist_sigma01.pdf}
%     \caption{Runtime}
%     \label{fig:inlatimehist}
% \end{figure}

\begin{table}[t]
\caption{Mean estimation runtime of latent GMRF model using log-likelihood, and $\LOOS$   for no outliers, and for data with one outlier in each of the 10 samples. Standard deviation is reported within parentheses.}
    \label{tab:latent:GMRF:time:hist}
    \centering
    \begin{tabular}{rcc}
    \toprule
       Scoring rule & 0 outliers & 10 outliers \\
       \hline 
       Log-likelihood & 3.61 (0.730)& 3.32 (0.760)\\
      $\LOOS$, $\slog$  &2.81 (0.598)&2.51 (0.570)\\
       $\LOOS$, SCRPS      & 2.75 (0.582)& 2.54 (0.553)\\
       $\LOOS$, $\sroot$ &2.58 (0.565)& 2.48 (0.509)\\
       $\LOOS$, CRPS       &  2.52 (0.520)&2.53 (0.596)\\
       $\LOOS$, rCRPS      & 2.77 (0.625) &2.75 (0.680 )\\
    \bottomrule
    \end{tabular}  
\end{table}

%Finally, Figure \ref{fig:godambe:m3}\todo{add figure} shows the asymptotic variances of the estimators based on the Godambe information matrices, and one can see that it... \todo{discuss}. 

% \begin{figure}
%     \centering    %\includegraphics{images/godambe/godambe_res_m220240630_103127_lines_dark2.pdf}
%     \caption{Asymptotic variance for the latent GMRF model using the inverse Godambe matrix with $\log\tau = ...$ (top) and $\log\kappa = ...$ (bottom).}
%     \label{fig:godambe:m3}
% \end{figure}

\subsection{A non-stationary model}
The final model we consider is a latent non-stationary model. We again model the data as 
$\Yb|\Xb \sim N(\Ab\Xb,\sigma_\varepsilon^2 I)$, where $\Xb \sim N(\mub, \Qb^{-1})$ now has  precision matrix  
$$
\Qb = \tau(\kappa^2 \Cb + \Gb) \Cb^{-1} (\kappa^2 \Cb + \Gb)\tau, 
$$
and where $\tau$ is a diagonal matrix with diagonal elements $\tau_i = \tau_0\sqrt{|s_{i1}|}$. This corresponds to introducing non-stationarity in $\tau$ of Eq. \eqref{eq:spde}, i.e. solving
\begin{equation*}\label{eq:spde:nonstationary}
    (\kappa^2-\Delta)^{\alpha/2}(\tau(s) \Xb)=\mathcal{W},
\end{equation*}
where $\tau(s) = \tau_0\sqrt{|s_{1}|}$.
The simulation setup is exactly the same as for the stationary latent GMRF model with $\thetab = (\sigma_\varepsilon,\tau_0,\kappa)= (0.5, 0.16, 1.75)$. 

\begin{figure}[t]
     \centering
     \includegraphics[width=\textwidth]{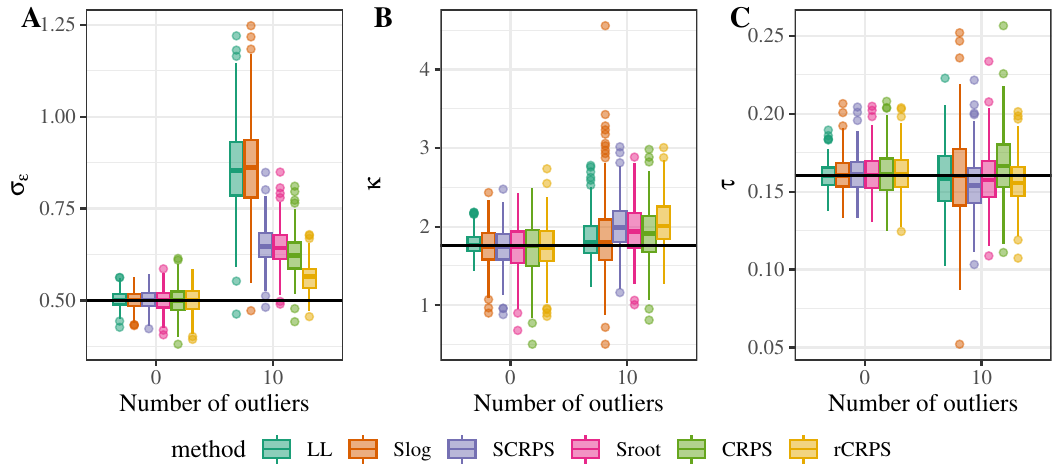}
     
     \caption{Estimated parameters for the non-stationary model based on data with and without outliers.}\label{fig:est:outlier:m4}
\end{figure}

The results can be seen in Figure~\ref{fig:est:outlier:m4}.
Without outliers, the variance of the estimates is smallest for the scale invariant scores, but largest for the CRPS. The $\sroot$ scores is less affected by the non-stationary scale of the data than the CRPS but still more than the logarithmic score and the SCRPS, and the variance of estimations reflects that. In the case of outliers, the results are similar to the previous results, where the rCRPS results in the most robust estimator.

\subsection{Predictive quality}

% \begin{figure}[t!]
%     \centering
%     \includegraphics[width=0.6\textwidth]{images/Ainlatesttrainoutliers/maternA_map_all.pdf}
%     \caption{Mesh along with observation points divided into testing and training points.}
%     \label{fig:test:train}
% \end{figure}

\begin{figure}[t]
    \centering
    \includegraphics[width=0.9\textwidth]{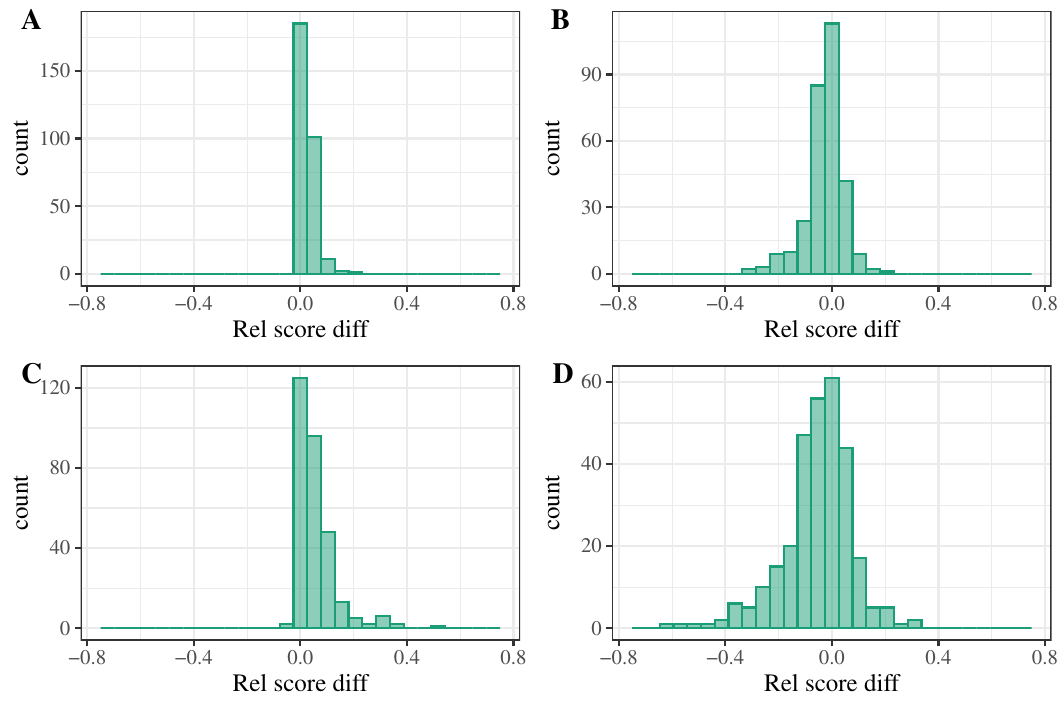}
    \caption{Proportional score difference using the estimated parameters from maximising $\LOOS_{\sroot}$ and Log-likelihood with positive values suggesting better predictive quality of $\LOOS$. The scores were computed on the training data (left), and on test data dependent on latent fields in training data (right), using $\sroot$ (top) and RMSE (bottom).}
    \label{fig:A:score}
\end{figure}

So far, we have only assessed the quality of the parameter estimates for the different estimators. However, as a common goal in spatial statistics is spatial prediction of the field at unobserved locations, we now consider the predictive quality of the models when the parameters are estimated using the different methods. Throughout the section, we focus on the latent GMRF model in Section~\ref{sec:latentGMRF}. 

We simulate observations at the training and test locations shown in Figure~\ref{fig:spdemesh} using the same parameters as in Section~\ref{sec:latentGMRF}. We then estimate the parameters using maximum likelihood and using the different LOOS estimators based on the training data. We then measure the predictive quality of the fitted models 
in two different ways:
\begin{enumerate}[(i)]
    \item By leave-one-out cross-validation on the training data. This is a common practice in spatial statistics, under the argument that leaving one point out of the training data will not affect the parameter estimate too much. 
    \item On the test data, with parameter estimates from the training data.
\end{enumerate}
This was repeated 300 times, each time for a different simulated dataset. The quality of predictions was measured through the average $\sroot$ score and through the root mean square error (RMSE). To compare the predictive quality, the relative score difference for $\hat{\theta}_{LL}$ from the log-likelihood and  $\hat{\theta}_{\sroot}$ from the $\sroot$ score was computed as 
\begin{equation*}
100\cdot\frac{S(\p_{\hat{\theta}_{\sroot}},\yb)-S(\p_{\hat{\theta}_{LL}},\yb)}{S(\p_{\hat{\theta}_{LL}},\yb)},
\end{equation*}
which corresponds to the percentual change in predictive quality when using $\LOOS$ estimator compared to the log-likelihood estimate. 

The results are shown in Figure \ref{fig:A:score}, where positive values suggest better prediction with $\sroot$ and negative suggest better prediction with log-likelihood. When the cross-validation is performed on the training dataset, the $\sroot$ approach outperforms the log-likelihood in prediction quality, which is expected as the LOOS estimator is optimising predictive quality. However, when considering predictions on the test data, there is essentially no difference between the methods. 

\begin{figure}[t]
    \centering
        \includegraphics[width=\textwidth]{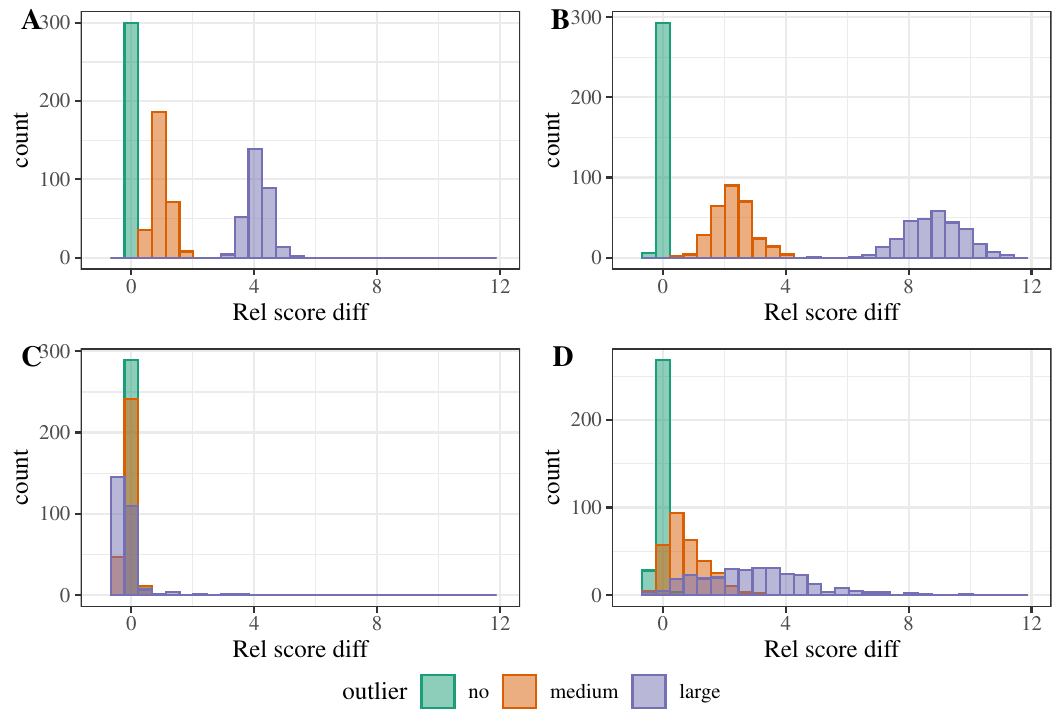}
    \caption{Proportional score difference using the estimated parameters from maximising $\LOOS_{\sroot}$ and Log-likelihood with different outlier values, with positive values suggesting better predictive quality of $\LOOS$. The scores were computed on the training data (left), and on test data dependent on latent fields in training data (right), using $\sroot$ (top) and RMSE (bottom). The outliers were set as $|y_i|+K$, with $K\in\{5,10\}$.}
    \label{fig:predictionoutlier}
\end{figure}

Next, we repeat the same procedure but when outliers are added to the training data. Again we replace $y_i$ by $|y_i|+K$ at a random location $i$, as done in Section \ref{sec:latentGMRF}, but we also investigate the effect of the outlier size and let $K\in\{5,10\}$ represent a medium and a large outlier. 
The results can be seen in Figure~\ref{fig:predictionoutlier}.
In this case, we can see a much bigger difference between the two methods. The predictive ability increases when using the LOOS estimator instead of maximum likelihood, and as the sizes of the outliers increase, the difference increases.

\section{Application to temperature reanalysis }\label{sec:results:casestudy}
\begin{figure}[t]
    \centering
    \includegraphics[width=0.7\textwidth]{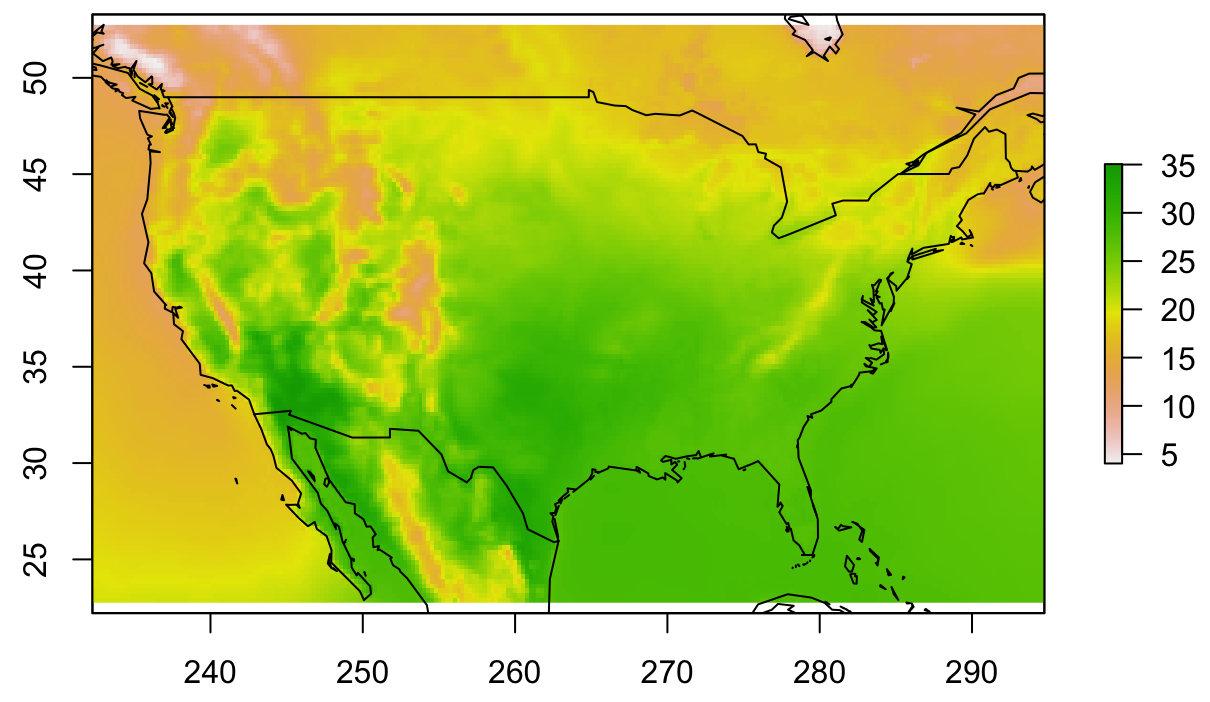}
    \caption{Average July temperature 1942.}
    \label{fig:tempdata}
\end{figure}

As an illustration of the methods, we consider statistical modelling of ERA5 temperature reanalysis data \citep{hersbach2020era5}. We consider the region shown in Figure~\ref{fig:tempdata} covering the contiguous United States and the the average July temperature at 2m above the surface of land, sea or inland waters for the years 1940 to 2023. As this is reanalysis data it is quite smooth, and we therefore model it as direct observations of the SPDE-based GMRF of Section~\ref{sec:results:gmrf} with $\alpha=2$. Specifically, if $\yb_t$ denotes the data at year $t$ we assume that this is an observation of $\Yb_t \sim N(\mub, \Qb^{-1})$, where $\Qb$ has the form \eqref{eq:spdeQ}, and the mesh is given by the regular longitude-latitude lattice on which the data is given. Further $\mub = \beta_0 + \beta_1\xb_1 + \beta_2\xb_2$, where $\xb_1$ is a vector with the geopotential height in kilometres at the observation locations and $\xb_2$ is a land-sea mask. Both of these covariates are available in the ERA5 data \citep{era5data}. 
We assume there is no temporal correlation, so that all 64 vectors $\Yb_t$ are independent and identically distributed. The dataset thus has 30371 spatial locations and 64 temporal replicates. 

We fit the model using maximum likelihood and using the LOOS estimator based on the $\sroot$ score. As before, the estimation is done using numerical maximisation using the \texttt{optim} function in R. The starting values for the parameters in optimisation are set as the ordinary least squares estimates for the regression coefficients, a value of $\kappa$ which corresponds to a practical correlation range of 5 degrees, and a value of $\tau$ that results in a variance of $\Yb$ that is the same as the variance of the residuals in the ordinary least squares estimation. 
The computation time for the parameter estimation was 1.53 hours for the maximum likelihood estimator and 49 seconds for the LOOS estimator. The parameter estimates can be seen in Table~\ref{tab:dataparam}. It can be noted that the estimated regression coefficients are similar between the two methods, but that the LOOS estimator gives a higher effect to elevation, whereas the maximum likelihood estimator gives a bigger effect to the land-sea mask. The main difference, however, is the estimation of $\tau$ and $\kappa$, where the lower $\kappa$ estimate and higher $\tau$ estimate of the LOOS estimator corresponds to a larger range and a smaller variance. 

\begin{table}[t]
\caption{Parameter estimates for the reanalysis data.}
    \label{tab:dataparam}
    \centering
    \begin{tabular}{rccccc}
    \toprule
       Estimation method & $\beta_0$ & $\beta_1$ & $\beta_2$ & $\kappa$ & $\tau$ \\
       \hline 
       maximum likelihood & 297.44 & -6.01  & 1.62 & 0.855 & 0.086 \\
       LOOS $\sroot$ & 295.79 & -6.71 & 1.32 & 0.002 & 0.197 \\
    \bottomrule
    \end{tabular}  
\end{table}

To check the goodness of fit, we perform leave-one-out cross-validation while keeping the parameter estimates fixed, as is standard in spatial statistics. The results can also be seen in Table~\ref{tab:dataparam}. Not surprisingly, the LOOS estimator has better predictive performance according to the $\sroot$ score, as this is what was optimised. However, as can be noted in the table, the LOOS estimator also provides better predictive performance in the CRPS, RMSE and SCRPS metrics. 
Thus, the LOOS estimator provided parameter estimates which are more than $100$ times faster to compute and give better predictive performance. 

\begin{table}[t]
\caption{Leave-one-out cross-validation metrics for the reanalysis data. All metrics are negatively oriented here, so that a lower value means a better score.}
    \label{tab:datacv}
    \centering
    \begin{tabular}{rcccc}
    \toprule
       Estimation method & $-\sroot$ & RMSE & $-$CRPS & $-$SCRPS  \\
       \hline 
       maximum likelihood   & 0.731 & 0.264 & 0.609 & 0.693 \\
       LOOS $\sroot$        & 0.655 & 0.216 & 0.607 & 0.575 \\
    \bottomrule
    \end{tabular}  
\end{table}

Clearly, the assumed model is not very realistic for the data, and it was mostly used to illustrate the computational efficiency of the LOOS estimator. Therefore, the predictive performance could likely be improved further by for example using a non-stationary model where $\kappa$ and $\tau$ are spatially varying \citep{Lindgren2011}, or by estimating the smoothness from data by using the rational SPDE approach of \cite{bolin2020rational, bolin2024covariance}. Temporal dependence could also be modelled through the non-separable SPDE models of \cite{lindgren2024diffusion}. Importantly, as all those models are Markov models, the computational efficiency of the LOOS estimator carries over to those models as well.

\section{Conclusion}\label{sec:conclusion}

In this article we have proposed a leave-one-out cross-validation score, $\LOOS$ as loss function for model inference on random field models in spatial statistics. The approach was studied for different choices of scoring rules in terms of statistical efficiency, time complexity, robustness and predictive performance. The scores used for the $\LOOS$ comparison were logarithmic score, CRPS, SCRPS and a new kernel score, $\sroot$, based on the generalised kernel score with properties between the scale invariant SCRPS and the non-scale invariant CRPS.

The $\LOOS$ estimators are asymptotically unbiased, as mentioned in \citet{Dawid2016MinimumInference}. Without outliers in the data, their variance is slightly larger than for the log-likelihood estimator. 
When the data contains outliers, the $\LOOS$ estimators with generalised kernel scores are more robust against outliers than the log-likelihood and LOOS estimation based on the log-score. Among the generalised kernel scores tested, the rCRPS gave the most robust estimates whereas the scale invariant SCRPS gave the lowest variance. The new $\sroot$ score provides a good balance between robustness and low asymptotic variance.

We showed how to compute the LOOS estimator efficiently for GMRFs. The main computational advantage of LOOS for such models is that it does not require the computation of the determinant of the precision matrix, which is the main computational cost in likelihood estimation. For latent GMRF models and for standard covariance-based Gaussian random fields, the computation time of the LOOS estimator is similar to that of maximum likelihood.

The LOOS estimator  can be computed without Monte Carlo approximation for each model where the score of conditional distributions is available in closed form. One interesting class of models to test LOOS estimation on in future work are the Type G non-Gaussian random fields discussed in \cite{bolin2020multivariate} and \cite{cabral2023controlling, cabral2024fitting}. These often has Markov properties, which means that the LOOS estimator can be computed with low computational cost. LOOS estimation is particularly interesting for such non-Gaussian models as likelihood-based estimation of flexible non-Gaussian models can be sensitive to outliers. 

Another interesting topic to investigate is to modify the score by using different types of cross-validation. One option, for example, would be to select a random subset of locations to predict in each iteration of the optimisation. That could greatly reduce the computational cost of inference for models without Markov properties.

%When the observations follow a multivariate normal distribution with a sparse precision matrix, the computations can be really fast for high dimensions since we do not need to compute the determinant of the matrix. When the precision matrix is not sparse, the method is of similar time-complexity as the log likelihood. The log likelihood provides more precise estimates, i.e. the estimates have lower variance. 

%\appendix
\begin{appendices}
\counterwithin*{figure}{section}
\renewcommand\thefigure{\thesection\arabic{figure}}  
\counterwithin*{equation}{section}
\renewcommand\theequation{\thesection\arabic{equation}}
\counterwithin*{table}{section}
\renewcommand\thetable{\thesection\arabic{table}}

\section{Local scale invariance of \texorpdfstring{$\sroot$}{Sroot}}\label{app:sroot:scale:invariance}
The concept of local scale invariance of proper score was defined by \citet{Bolin2023} as a means of describing scores that are not affected by the magnitude of predictive uncertainty. If $\p_{\thetab}$ is a location-scale family with parameters $\thetab=(\mu,\sigma)$, such as a univariate Gaussian distribution, and one considers a small perturbation $d\thetab$ of the parameters, then locally one can write
$$
S(\p_{\thetab},\p_{\thetab})-S(\p_{\thetab+d\thetab},\p_{\thetab})=d\thetab^\top s(\p_{\thetab})d\thetab
$$
since $S$ is proper. 
The function $s(\p_{\thetab})$ is the scale function of $S$ on $\p$. If the scale function exists and can be written as
$$s(\p_{\thetab})=\frac{1}{\sigma^2}s(\p_{0,1})$$
then $S$ is said to be locally scale invariant. The important thing here is the scaling $1/\sigma^2$ with the scale, which is what makes the magnitude of the score invariant to the magnitude of the prediction uncertainty. The dependence on $\sigma$ for the scale function of different scoring rules is shown in Table~\ref{tab:summaryscores}, and there one can see that the log-score and the SCRPS are scale invariant.
For further details, see \citet{Bolin2023}. 

We now derive the scale function of $\sroot$, which has not been investigated before. Let $\p_\theta$ be a prediction with parameters $\theta$. For shorter notation, we define $G_{\p,\Q}=E_{\p,\Q}[|X-Y|]$ and $G_\p=G_{\p,\p}$. In addition, we have the following notation for their derivatives: $\E_{\dot{P}Q}=\nabla_{\thetab}G_{\p_{\thetab},\Q}$, $\E_{Q\dot{P}}=\nabla_{\thetab}G_{\Q,\p_{\thetab}}$, and $\E_{\ddot{P}Q}=\nabla_{\thetab}^2G_{\p_{\thetab},\Q}$. The $\E_{\dot{P}\dot{P}}$ denotes the gradient applied to both arguments of $G_{\p_{\thetab},\p_{\thetab}}$. Consider the score derivatives wrt $\theta$, i.e. $\nabla_{\theta}\sroot(\p_\theta,\Q)$ and $\nabla^2_{\theta}\sroot(\p_\theta,\Q)|_{\Q=\p}$. Firstly,
\begin{equation*}
    \begin{aligned}
        \nabla_{\theta}\sroot(\p_\theta,\Q)=\frac{\E_{\dot{P}Q}}{\sqrt{G_{\p_\theta}}}-\frac{G_{\p_\theta,\Q}\E_{\dot{P}P}}{G_{\p_\theta}^{3/2}}
    \end{aligned}
\end{equation*}
and
\begin{equation*}
    \begin{aligned}
    \nabla^2_{\theta}\sroot(\p_\theta,\Q)=\frac{\E_{\ddot{P}Q}}{\sqrt{G_{\p_\theta}}}-\frac{\E_{\dot{P}Q}\E_{\dot{P}P}^\top}{G_{\p_\theta}^{3/2}}-\frac{\E_{\dot{P}Q}\E_{\dot{P}P}^\top}{G_{\p_\theta}^{3/2}}-\frac{G_{\p_\theta,\Q}(\E_{\ddot{P}P}+\E_{\dot{P}\dot{P}})}{G_{\p_\theta}^{3/2}}+\frac{3G_{\p_\theta,\Q}\E_{\dot{P}P}\E_{\dot{P}P}}{G_{\p_\theta}^{5/2}}.
    \end{aligned}
\end{equation*}
Inserting $\Q=\p_\theta$ gives
\begin{equation*}
    \begin{aligned}       \nabla^2_{\theta}\sroot(\p_\theta,\Q)|_{\Q=\p_\theta}&=\frac{\E_{\ddot{P}P}}{\sqrt{G_{\p_\theta}}}-\frac{\E_{\dot{P}P}\E_{\dot{P}P}^\top}{G_{\p_\theta}^{3/2}}-\frac{\E_{\dot{P}P}\E_{\dot{P}P}^\top}{G_{\p_\theta}^{3/2}}-\frac{\E_{\ddot{P}P}+\E_{\dot{P}\dot{P}}}{\sqrt{G_{\p_\theta}}}+\frac{3\E_{\dot{P}P}\E_{\dot{P}P}}{G_{\p_\theta}^{3/2}}\\
        &=\frac{\E_{\dot{P}P}\E_{\dot{P}P}^\top}{G_{\p_\theta}^{3/2}}-\frac{\E_{\dot{P}\dot{P}}}{\sqrt{G_{\p_\theta}}}\\
         &=\frac{1}{\sqrt{G_{\p_\theta}}}\left(\frac{\E_{\dot{P}P}\E_{\dot{P}P}^\top}{G_{\p_\theta}}-\E_{\dot{P}\dot{P}}\right).
    \end{aligned}
\end{equation*}
This gives scale function
\begin{equation*}
    s(\Q_\theta)=\frac{1}{\sigma\sqrt{\sigma}}\E_{\Q,\Q}[H_\Q(X,Y)]
\end{equation*}
where $H_\Q$ is a $2\times2$ positive semi-definite matrix independent of $\theta$. The scoring rule $\sroot$ is therefore not locally scale invariant as the logarithmic score and the SCRPS, but it is less dependent on scale than the CRPS and the Hyvärinen score. According to \citet{Bolin2023}, the scoring rule is then also homogeneous of order $1/2$, which should not be confused with homogeneity in the density.

\section{Score equations for the Normal distribution}\label{app:normal:score}

For $\p=N(\mu,\sigma^2)$, we have

\begin{equation*}
    \E_\p[|X-y|]=2\sigma\phi\left(\frac{\mu-y}{\sigma}\right)+(\mu-y)\left(2\Phi\left(\frac{\mu-y}{\sigma}\right)-1\right)
\end{equation*}
and
    $\E_{\p,\p}[|X-X'|]=2\frac{\sigma}{\sqrt{\pi}}$,
where $\phi$ and $\Phi$ are the density and distribution function of the standard normal distribution, respectively.
This gives

\begin{equation*}\label{eq:crps:normal}
\begin{aligned}
    \CRPS(\p,y)&=\frac{1}{2}\E_{\p\p}[|X-X'|]-\E_\p[|X-y|]\\
    &=\frac{\sigma}{\sqrt{\pi}}-2\sigma\phi\left(\frac{y-\mu}{\sigma}\right)-(y-\mu)\left(2\Phi\left(\frac{y-\mu}{\sigma}\right)-1\right),\\
    \sroot(\p,y)&=-\frac{\E_\p}{\sqrt{\E_{\p,\p}}}
    =-\frac{2\sigma\phi\left(\frac{y-\mu}{\sigma}\right)+(y-\mu)\left(2\Phi\left(\frac{y-\mu}{\sigma}\right)-1\right)}{\sqrt{2\sigma/\sqrt{\pi}}},\\
    \SCRPS(\p,y)&=-\frac{\E_\p}{\E_{\p\p}}-\frac{1}{2}\log \E_{\p\p}\\
    &=-\sqrt{\pi}\phi\left(\frac{y-\mu}{\sigma}\right)-\frac{\sqrt{\pi}(y-\mu)}{2\sigma}\left(2\Phi\left(\frac{y-\mu}{\sigma}\right)-1\right)-\frac{1}{2}\log (2\sigma/\sqrt{\pi}).
    \end{aligned}
\end{equation*}
Moreover, the robust CRPS (rCRPS) with kernel function $g(x,y)=|x-y|1\{|x-y|<c\}$ is of the form \citep{Bolin2023}
\begin{equation*}
    \rCRPS(\p,y)=0.5h(0,\sqrt{2}\sigma,c)-h(\mu-y,\sigma,c)
\end{equation*}
with
\begin{equation*}
\begin{aligned}
    h(\mu,\sigma,c)=&\sigma\left(2\phi\left(\frac{\mu}{\sigma}\right)-\phi\left(\frac{c-\mu}{\sigma}\right)-\phi\left(\frac{c+\mu}{\sigma}\right)\right)-\mu\\
    &+(c-\mu)\Phi\left(\frac{\mu-c}{\sigma}\right)+2\mu\Phi\left(\frac{\mu}{\sigma}\right)+(\mu+c)\Phi\left(\frac{-c-\mu}{\sigma}\right).
\end{aligned}
\end{equation*}
\end{appendices}

\bibliography{bibfin,sample, ref2}

\begin{thebibliography}{32}
\providecommand{\natexlab}[1]{#1}
\providecommand{\url}[1]{\texttt{#1}}
\expandafter\ifx\csname urlstyle\endcsname\relax
  \providecommand{\doi}[1]{doi: #1}\else
  \providecommand{\doi}{doi: \begingroup \urlstyle{rm}\Url}\fi

\bibitem[Baringhaus and Franz(2004)]{Baringhaus2004}
L.~Baringhaus and C.~Franz.
\newblock {On a new multivariate two-sample test}.
\newblock \emph{Journal of Multivariate Analysis}, 88\penalty0 (1):\penalty0 190--206, 2004.
\newblock \doi{10.1016/S0047-259X(03)00079-4}.

\bibitem[Besag(1975)]{Besag1975StatisticalData}
J.~Besag.
\newblock {Statistical Analysis of Non-Lattice Data}.
\newblock \emph{Journal of the Royal Statistical Society. Series D (The Statistician)}, 24\penalty0 (3):\penalty0 179--195, 1975.
\newblock \doi{https://doi.org/10.2307/2987782}.

\bibitem[Bolin and Kirchner(2020)]{bolin2020rational}
D.~Bolin and K.~Kirchner.
\newblock {The rational SPDE approach for {G}aussian random fields with general smoothness}.
\newblock \emph{Journal of Computational and Graphical Statistics}, 29\penalty0 (2):\penalty0 274--285, 2020.

\bibitem[Bolin and Wallin(2020)]{bolin2020multivariate}
D.~Bolin and J.~Wallin.
\newblock {Multivariate type G Mat{\'e}rn stochastic partial differential equation random fields}.
\newblock \emph{Journal of the Royal Statistical Society Series B: Statistical Methodology}, 82\penalty0 (1):\penalty0 215--239, 2020.

\bibitem[Bolin and Wallin(2023)]{Bolin2023}
D.~Bolin and J.~Wallin.
\newblock {Local scale invariance and robustness of proper scoring rules}.
\newblock \emph{Statistical Science}, 38\penalty0 (1):\penalty0 140--159, 2023.
\newblock ISSN 21688745.
\newblock \doi{10.1214/22-STS864}.

\bibitem[Bolin et~al.(2024)Bolin, Simas, and Xiong]{bolin2024covariance}
D.~Bolin, A.~B. Simas, and Z.~Xiong.
\newblock {Covariance--based rational approximations of fractional {SPDEs} for computationally efficient {B}ayesian inference}.
\newblock \emph{Journal of Computational and Graphical Statistics}, 33\penalty0 (1):\penalty0 64--74, 2024.

\bibitem[Cabral et~al.(2023)Cabral, Bolin, and Rue]{cabral2023controlling}
R.~Cabral, D.~Bolin, and H.~Rue.
\newblock {Controlling the flexibility of non-Gaussian processes through shrinkage priors}.
\newblock \emph{Bayesian Analysis}, 18\penalty0 (4):\penalty0 1223--1246, 2023.

\bibitem[Cabral et~al.(2024)Cabral, Bolin, and Rue]{cabral2024fitting}
R.~Cabral, D.~Bolin, and H.~Rue.
\newblock {Fitting latent non-Gaussian models using variational Bayes and Laplace approximations}.
\newblock \emph{Journal of the American Statistical Association}, pages 1--13, 2024.

\bibitem[Carvalho(2016)]{Carvalho2016AnRules}
A.~Carvalho.
\newblock {An overview of applications of proper scoring rules}.
\newblock \emph{Decision Analysis}, 13\penalty0 (4):\penalty0 223--242, 2016.
\newblock ISSN 15458504.
\newblock \doi{10.1287/deca.2016.0337}.

\bibitem[Dawid(2007)]{Dawid2007TheRules}
A.~P. Dawid.
\newblock {The geometry of proper scoring rules}.
\newblock \emph{Annals of the Institute of Statistical Mathematics}, 59\penalty0 (1):\penalty0 77--93, 3 2007.
\newblock \doi{10.1007/s10463-006-0099-8}.

\bibitem[Dawid et~al.(2016)Dawid, Musio, and Ventura]{Dawid2016MinimumInference}
A.~P. Dawid, M.~Musio, and L.~Ventura.
\newblock {Minimum Scoring Rule Inference}.
\newblock \emph{Scandinavian Journal of Statistics}, 43\penalty0 (1):\penalty0 123--138, 2016.
\newblock ISSN 14679469.
\newblock \doi{10.1111/sjos.12168}.

\bibitem[De~Fondeville and Davison(2018)]{DeFondeville2018}
R.~De~Fondeville and A.~C. Davison.
\newblock {High-dimensional peaks-over-threshold inference}.
\newblock \emph{Biometrika}, 105\penalty0 (3):\penalty0 575--592, 2018.
\newblock ISSN 14643510.
\newblock \doi{10.1093/biomet/asy026}.

\bibitem[Friederichs and Thorarinsdottir(2012)]{Friederichs2012}
P.~Friederichs and T.~L. Thorarinsdottir.
\newblock {Forecast verification for extreme value distributions with an application to probabilistic peak wind prediction}.
\newblock \emph{Environmetrics}, 23\penalty0 (7):\penalty0 579–594, 4 2012.
\newblock \doi{10.1002/env.2176}.

\bibitem[Gneiting and Raftery(2007)]{Gneiting2007}
T.~Gneiting and A.~E. Raftery.
\newblock {Strictly proper scoring rules, prediction, and estimation}.
\newblock \emph{Journal of the American Statistical Association}, 102\penalty0 (477):\penalty0 359--378, 3 2007.
\newblock \doi{10.1198/016214506000001437}.

\bibitem[Gneiting et~al.(2005)Gneiting, Raftery, Westveld, and Goldman]{Gneiting2005CalibratedEstimation}
T.~Gneiting, A.~E. Raftery, A.~H. Westveld, and T.~Goldman.
\newblock {Calibrated probabilistic forecasting using ensemble model output statistics and minimum CRPS estimation}.
\newblock \emph{Monthly Weather Review}, 133\penalty0 (5):\penalty0 1098–1118, 2005.
\newblock ISSN 00270644.
\newblock \doi{10.1175/MWR2904.1}.

\bibitem[Gneiting et~al.(2008)Gneiting, Stanberry, Grimit, Held, and Johnson]{Gneiting2008}
T.~Gneiting, L.~I. Stanberry, E.~P. Grimit, L.~Held, and N.~A. Johnson.
\newblock {Assessing probabilistic forecasts of multivariate quantities, with an application to ensemble predictions of surface winds}.
\newblock \emph{Test}, 17\penalty0 (2):\penalty0 211--235, 2008.
\newblock ISSN 11330686.
\newblock \doi{10.1007/s11749-008-0114-x}.

\bibitem[Godambe(1960)]{Godambe1960}
V.~P. Godambe.
\newblock {An optimum property of regular maximum likelihood estimation}.
\newblock \emph{The Annals of Mathematical Statistics}, 31\penalty0 (4):\penalty0 1208 -- 1211, 1960.
\newblock \doi{10.1214/aoms/117770569}.
\newblock URL \url{https://doi.org/10.1214/aoms/1177705693}.

\bibitem[Good(1952)]{Good1952RationalDecisions}
I.~J. Good.
\newblock {Rational Decisions}.
\newblock \emph{Journal of the Royal Statistical Society: Series B (Methodological)}, 14\penalty0 (1):\penalty0 107–114, 1952.
\newblock ISSN 1369-7412.
\newblock \doi{10.1111/j.2517-6161.1952.tb00104.x}.

\bibitem[Hampel(1974)]{Hampelincluencefunc}
F.~R. Hampel.
\newblock The influence curve and its role in robust estimation.
\newblock \emph{J. Amer. Statist. Assoc.}, 69:\penalty0 383--393, 1974.
\newblock ISSN 0162-1459,1537-274X.
\newblock URL \url{http://links.jstor.org/sici?sici=0162-1459(197406)69:346<383:TICAIR>2.0.CO;2-N&origin=MSN}.

\bibitem[Hersbach et~al.(2020)Hersbach, Bell, Berrisford, Hirahara, Hor{\'a}nyi, Mu{\~n}oz-Sabater, Nicolas, Peubey, Radu, Schepers, et~al.]{hersbach2020era5}
H.~Hersbach, B.~Bell, P.~Berrisford, S.~Hirahara, A.~Hor{\'a}nyi, J.~Mu{\~n}oz-Sabater, J.~Nicolas, C.~Peubey, R.~Radu, D.~Schepers, et~al.
\newblock {The ERA5 global reanalysis}.
\newblock \emph{Quarterly Journal of the Royal Meteorological Society}, 146\penalty0 (730):\penalty0 1999--2049, 2020.

\bibitem[{Hersbach, H. and Bell, B. and Berrisford, P. and Biavati, G. and Hor\'anyi, A. and Munoz Sabater, J. and Nicolas, J. and Peubey, C. and Radu, R. and Rozum, I. and Schepers, D. and Simmons, A. and Soci, C. and Dee, D. and Th\'epaut, J-N.}(2018)]{era5data}
{Hersbach, H. and Bell, B. and Berrisford, P. and Biavati, G. and Hor\'anyi, A. and Munoz Sabater, J. and Nicolas, J. and Peubey, C. and Radu, R. and Rozum, I. and Schepers, D. and Simmons, A. and Soci, C. and Dee, D. and Th\'epaut, J-N.}
\newblock Era5 hourly data on single levels from 1940 to present. copernicus climate change service (c3s) climate data store (cds), 2018.
\newblock Accessed on 02-JUL-2024.

\bibitem[Huber and Ronchetti(2011)]{huber2011robust}
P.~J. Huber and E.~M. Ronchetti.
\newblock \emph{Robust statistics}.
\newblock John Wiley \& Sons, 2011.

\bibitem[Hyv{\"{a}}rinen(2005)]{Hyvarinen2005EstimationMatching}
A.~Hyv{\"{a}}rinen.
\newblock {Estimation of Non-Normalized Statistical Models by Score Matching}.
\newblock \emph{Journal of Machine Learning Research}, 6:\penalty0 1, 2005.

\bibitem[Hyv{\"{a}}rinen(2007)]{Hyvarinen2007SomeMatching}
A.~Hyv{\"{a}}rinen.
\newblock {Some extensions of score matching}.
\newblock \emph{Computational Statistics and Data Analysis}, 51\penalty0 (5):\penalty0 2499--2512, 2007.
\newblock ISSN 01679473.
\newblock \doi{10.1016/j.csda.2006.09.003}.

\bibitem[Lerch and Thorarinsdottir(2013)]{Lerch2013ComparisonForecasting}
S.~Lerch and T.~L. Thorarinsdottir.
\newblock {Comparison of non-homogeneous regression models for probabilistic wind speed forecasting}.
\newblock \emph{Tellus A: Dynamic Meteorology and Oceanography}, 65\penalty0 (1), 2013.
\newblock ISSN 0280-6495.
\newblock \doi{10.3402/tellusa.v65i0.21206}.

\bibitem[Lindgren et~al.(2011)Lindgren, Rue, and Lindstr{\"{o}}m]{Lindgren2011}
F.~Lindgren, H.~Rue, and J.~Lindstr{\"{o}}m.
\newblock {An explicit link between Gaussian fields and Gaussian Markov random fields: the stochastic partial differential equation approach}.
\newblock \emph{Journal of the Royal Statistical Society, Series B}, 73\penalty0 (4):\penalty0 423--498, 2011.
\newblock ISSN 13697412.
\newblock \doi{10.1111/j.1467-9868.2011.00777.x}.
\newblock URL \url{http://dx.doi.org/10.1111/j.1467-9868.2011.00777.x}.

\bibitem[Lindgren et~al.(2024)Lindgren, Bakka, Bolin, Krainski, and Rue]{lindgren2024diffusion}
F.~Lindgren, H.~Bakka, D.~Bolin, E.~Krainski, and H.~Rue.
\newblock {A diffusion-based spatio-temporal extension of {G}aussian {M}at{\'e}rn fields}.
\newblock \emph{SORT-Statistics and Operations Research Transactions}, pages 3--66, 2024.

\bibitem[{R Core Team}(2021)]{Rsoftware}
{R Core Team}.
\newblock \emph{{R: A Language and Environment for Statistical Computing}}.
\newblock R Foundation for Statistical Computing, Vienna, Austria, 2021.
\newblock URL \url{https://www.R-project.org/}.

\bibitem[Rue and Held(2005)]{Rue2005}
H.~Rue and L.~Held.
\newblock \emph{{Gaussian markov random fields: Theory and applications}}.
\newblock Chapman and Hall/CRC, 2005.
\newblock ISBN 9780203492024.
\newblock \doi{10.1198/tech.2006.s352}.

\bibitem[Rue et~al.(2009)Rue, Martino, and Chopin]{Rue2009}
H.~Rue, S.~Martino, and N.~Chopin.
\newblock {Approximate Bayesian inference for latent Gaussian models by using integrated nested Laplace approximations}.
\newblock \emph{Journal of the Royal Statistical Society. Series B: Statistical Methodology}, 71\penalty0 (2):\penalty0 319--392, 2009.
\newblock ISSN 13697412.
\newblock \doi{10.1111/j.1467-9868.2008.00700.x}.

\bibitem[Serafini et~al.(2022)Serafini, Naylor, Lindgren, Werner, and Main]{Serafini2022RankingEnvironment2}
F.~Serafini, M.~Naylor, F.~Lindgren, M.~J. Werner, and I.~Main.
\newblock Ranking earthquake forecasts using proper scoring rules: binary events in a low probability environment.
\newblock \emph{Geophysical Journal International}, 230\penalty0 (2):\penalty0 1419--1440, 2022.

\bibitem[Zamo and Naveau(2018)]{Zamo2018}
M.~Zamo and P.~Naveau.
\newblock {Estimation of the Continuous Ranked Probability Score with Limited Information and Applications to Ensemble Weather Forecasts}.
\newblock \emph{Mathematical Geosciences}, 50\penalty0 (2):\penalty0 209--234, 2 2018.
\newblock \doi{10.1007/s11004-017-9709-7}.

\end{thebibliography}

\end{document}